\documentstyle[12pt]{article}
\begin{document}

\title{Spinning Particle Dynamics \newline
on Six-Dimensional Minkowski Space}
\author{S. L. Lyakhovich, A. A. Sharapov and K. M. Shekhter}
\date{{\it Department of Physics, Tomsk State University,}\\
{\it Lenin Ave. 36, Tomsk 634050, Russia }}
\maketitle

\begin{abstract}
Massive spinning particle in $6d$-Minkowski space is described as a
mechanical system with the configuration space ${\ R}^{5,1}\times {\ CP}^3$.
The action functional of the model is unambiguously determined by the
requirement of identical (off-shell) conservation for the phase-space
counterparts of three Casimir operators of Poincar\'e group. The model
proves to be completely solvable. Its generalization to the constant
curvature background is presented. Canonical quantization of the theory
leads to the relativistic wave equations for the irreducible $6d$ fields.
\end{abstract}
\newpage
\section{Introduction}

A classical description of relativistic spinning particles is one of
the traditional branches of theoretical physics having a long history \cite
{c1,c2,c3}. By now, several approaches to this problem have been developed.
Most of the researches are based on the enlargement of the Minkowski space
by extra variables, anticommuting \cite{c2} or commuting \cite{c3,c13,c16},
responsible for the spin evolution. Being well adapted for the quantization,
the theories using Grassmann variables encounter, however, difficulties on
attempting to justify them at the classical level. Besides that, the
quantization of these theories lead to the Poincar\'e representation of
fixed spin.

The orbit method, developed in \cite{kir}, is the universal approach for the
description of the elementary systems. The basic
object of this approach is a presymplectic manifold ${\cal E}$, being a
homogeneous transformation space for a certain Lie group $G$, and the system
is considered as "elementary" for this group. The manifold carries
the invariant and degenerate closed
two-form $\Omega $ such that quotient space ${\cal E}/\ker \Omega $ is a
homogeneous symplectic manifold (in fact, it may be identified with some
covering space for coadjoint orbit ${\cal O}$ of group $G$). If $\theta $ is
a potential one--form for $\Omega $ then the first-order action functional
of the system may be written as%
$$
S=\int \theta
$$
Being applied to the Poincar\'e group, this method gives the Souriau
classification of the spinning particles. Meanwhile, there is another
trend to describe a spinning particle by means of a traditional
formalism based on an appropriate choice of the configuration space for
spin [1-5].

In a recent paper \cite{c13}, the new model was proposed for a massive
particle of arbitrary spin in $d=4$ Minkowski space to be a mechanical
system with the configuration space $R^{3,1}\times S^2$, where two sphere $%
S^2$ corresponds to the spinning degrees of freedom. It was shown that
principles underlying the model have simple physical and geometrical origin.
Quantization of the model leads to the unitary massive representations of
the Poincar\'e group. The model allows the direct extension to the case of
higher superspin superparticle and the generalization to the anti-de Sitter
space.

Despite the apparent simplicity of the model's construction, its higher
dimensional generalization is not so evident, and the most crucial point is
the choice of configuration space for spin. In this work we describe the
massive spinning particle in six-dimensional Minkowski space $R^{5,1}$, that
may be considered as a first step towards the uniform model construction for
all higher dimensions. It should also be noted that this generalization may
have a certain interest in its own rights since six is the one in every four
dimensions: 3, 4, 6 and 10 possess the remarkable properties such as the
presence of two-component spinor formalism or light-likeness of the spinor
bilinear \cite{ced}. These properties are conditioned by the connection
between the division algebras and the Lorentz groups of these spaces \cite
{kts}%
$$
SL\left( 2,A\right) \sim SO^{\uparrow }\left( \dim A+1,1\right)
$$
where A are the division algebras $R,C,H,O$ of real and complex numbers,
quaternions and octonions respectively. Besides that, these are exactly the
dimensions where the classical theory of Green-Schwarz superstring can be
formulated \cite{gs}.

Let us now sketch the broad outlines of the construction. First of all, for
any even dimension $d$, the model's configuration space is chosen to be the
direct product of Minkowski space $R^{d-1,1}$ and some $m$-dimensional
compact manifold $K^m$ being a homogeneous transformation space for the
Lorentz group $SO(d-1,1)$. Then the manifold $M^{d+m}=R^{d-1,1}\times K^m$
proves to be the homogeneous transformation space for the Poincar\'e group.
The action of the Poincar\'e group on $M^{d+m}$ is unambiguously lifted up to
the action on the cotangent bundle $T^{*}(M^{d+m})$ being the extended phase
space of the model. It is well known that the massive unitary irreducible
representations of the Poincar\'e group are uniquely characterized by the
eigenvalues of $d/2$ Casimir operators%
$$
C_1={\bf P}^2\ ,\ C_{i+1}={\bf W}^{A_1...A_{2i-1}}{\bf W}_{A_1...A_{2i-1}}\
,\quad i=1,...,\frac{d-2}2\ ,
$$
where{\bf \ }${\bf W}_{A_1....A_{2i-1}}=\epsilon _{A_1...A_d}{\bf J}%
^{A_{2i}A_{2i+1}}...{\bf J}^{A_{d-2}A_{d-1}}{\bf P}^{A_d}$ and ${\bf J}_{AB},%
{\bf P}_C$ are the Poincar\'e generators. This leads us to require the
identical (off-shell) conservation for the quantum numbers associated with
the phase space counterparts of Casimir operators. In other words $d/2$
first-class constraints should appear in the theory.

Finally, the dimensionality $m$ of the manifold $K^m$ is specified from the
condition that the reduced (physical) phase space of the model should be a
homogeneous symplectic manifold of Poincar\'e group (in fact, it should
coincide with the coadjoint orbit of maximal dimension $d^2/2$). A simple
calculation leads to $m=d(d-2)/4$. In the case of four-dimensional Minkowski
space this yields $m=2$ and two-sphere $S^2$ turns out to be the unique
candidate for the internal space of the spinning degrees of freedom. In the
case considered in this paper $d=6$, and hence $m=6$. As will be shown below
the suggestive choice for $K^6$ is the complex projective space $CP^3$.

The models can be covariantly quantized a l\^a Dirac by imposing the
first-class constraints on the physical states being the smooth complex
functions on the homogeneous space $M^{d(d+2)/4}=R^{d-1,1}\times
K^{d(d-2)/4} $%
$$
(\widehat{C}_i-\delta _i)\Psi =0\quad ,\qquad i=1,...,\frac d2\ ,
$$
where the parameters $\delta _i$ are the quantum numbers characterizing the
massive unitary representation of the Poincar\'e group. Thus the
quantization of the spinning particle theories reduces to the standard
mathematical problem of harmonic analysis on homogeneous spaces. It should
be remarked that manifold $M^{d(d+2)/4}$ may be thought of as the ${\it %
minimal}$ (in sense of its dimensionality) one admitting a non-trivial
dynamics of arbitrary spin, and hence it is natural to expect that the
corresponding Hilbert space of physical states will carry the ${\it %
irreducible}$ representation of the Poincar\'e group.

The paper is organized as follows. Sec.2 deals with the description of the
configuration space geometry, its local structure and various
parametrizations. In sec.3 we derive the model's action functional in the
first order formalism. We also consider the solutions of classical
equations of motion and discuss the geometry of the trajectories.
In sec.4 the second order
formalism for the theory is presented and the different reduced forms of
Lagrangian are discussed. Here we also investigate the causality conditions
for the theory. Sec.5 is devoted to the quantization of the theory in the
Hilbert space of smooth tensor fields over $M^{12}$. The connection with the
relativistic wave equations is apparently stated. In the conclusion we
discuss the received results and further perspectives. In the Appendix we have
collected the basic facts of half spinorial formalism in six-dimensions.

\section{Geometry of the configuration space \newline
and covariant parametrization}

We start with describing a covariant realization for the model's
configuration space chosen as $M^{12}=R^{5,1}\times CP^3.$ The manifold $%
M^{12}$ is the homogeneous transformation space for Poincar\'e group $P$
and, hence, it can be realized as a coset space $P/H$ for some subgroup $%
H\subset P.$ In order to present the subgroup $H$ in an explicit form it is
convenient to make Iwasawa decomposition for six-dimensional Lorentz group $%
SO\left( 5,1\right) $ in maximal compact subgroup $SO\left( 5\right) $ and
solvable factor $R$%
\begin{equation}
\label{a}SO\left( 5,1\right) =SO\left( 5\right) R
\end{equation}
Then the minimal parabolic subgroup, being defined as normalizer of $R$ in $%
SO\left( 5,1\right) $, coincides with $SO\left( 4\right) R.$ By means of the
decomposition $SO\left( 4\right) =SO\left( 3\right) \times SO\left( 3\right)
$ the subgroup $H$ is identified with $\left[ SO\left( 2\right) \times
SO\left( 3\right) \right] R$. Thus
\begin{equation}
\label{b}M^{12}=R^{5,1}\times CP^3\sim \frac{Poincar\acute e\,group}{\left[
SO\left( 2\right) \times SO\left( 3\right) \right] R}\sim R^{5,1}\times
\frac{SO\left( 5\right) }{SO\left( 2\right) \times SO\left( 3\right) }
\end{equation}
and thereby one has the isomorphism
\begin{equation}
\label{c}CP^3\sim \frac{SO\left( 5\right) }{SO\left( 2\right) \times
SO\left( 3\right) }
\end{equation}
Furthermore, from the sequence of the subgroups
\begin{equation}
\label{d}SO\left( 2\right) \times SO\left( 3\right) \subset SO\left(
4\right) \subset SO\left( 5\right)
\end{equation}
and the obvious isomorphisms $S^4\sim SO\left( 5\right) /SO\left( 4\right)
,S^2\sim SO\left( 3\right) /SO\left( 2\right) $ one concludes that $CP^3$
may be considered as the bundle $CP^3\rightarrow S^4$ with the fibre $S^2$%
. The fibres lie in $CP^3$ as projective lines $CP^1\sim S^2$. Thus, $CP^3$
is locally represented as
\begin{equation}
\label{f}CP^3\stackrel{loc.}{\sim }S^4\times S^2
\end{equation}
Note that the subgroup $H$ contains solvable factor ${\cal R}$ (and
hence $H$ is not an unimodular), so there is no Poincar\'e invariant
measure on $M^{12}$. Nevertheless from rel.(\ref{c}) it follows that there is a
quasi-invariant measure which becomes a genuine invariant when the Lorentz
transformations are restricted to the stability subgroup of time-like vector
$ SO\left( 5\right) $.

In spite of the quite intricate structure, the subgroup $H$ admits a simple
realization, namely, it can be identified with all the $SO\left( 5,1\right)
- $transformations multiplying the Weyl spinor $\lambda $ by a complex
factor
\begin{equation}
\label{g}N_a{}^b\lambda _b=\alpha \lambda _a\quad ,\quad \alpha \in
C\backslash \left\{ 0\right\}
\end{equation}
(all the details concerning six-dimensional spinor formalism are collected
in the Appendix). This observation readily leads to the covariant
parametrization of $CP^3$ by a complex Weyl spinor subject to the equivalence
relation
\begin{equation}
\label{h}\lambda _a\sim \alpha \lambda _a\qquad ,\qquad \alpha \in
C\backslash \left\{ 0\right\}
\end{equation}

By construction, the Poincar\'e group generators act on $M^{12}$ by the
following vector fields:
\begin{equation}
\label{i}{\bf P}^A=\partial ^A\quad ,\quad {\bf M}_{AB}=x_A\partial
_B-x_B\partial _A-\left( \left( \sigma _{AB}\right) _a{}^b\lambda _b\partial
^a~+~c.c.\right)
\end{equation}
where $\left\{ x^A\right\} $ are the Cartesian coordinates on $R^{5,1}$. It
is evident that Poincar\'e generators commute with the projective
transformations (\ref{h}) generated by the vector fields
\begin{equation}
\label{j}d=\lambda _a\partial ^a\qquad ,\qquad \overline{d}=\overline{%
\lambda }_a\overline{\partial }^a
\end{equation}
Then the space of scalar functions on $M^{12}$ is naturally identified with
those functions $\Phi \left( x^A,\lambda _a,\overline{\lambda }_b\right) $
which satisfy the homogeneity conditions
\begin{equation}
\label{k}d\Phi =\overline{d}\Phi =0
\end{equation}
Let us consider the ring of invariant differential operators acting on the
space of scalar functions on $M^{12}$. Such operators should commute with
the Poincar\'e transformations (\ref{i}) and the projective ones (\ref{j}).
It is easy to see that there are only three independent Laplace operators.
They are
\begin{equation}
\label{l}
\begin{array}{c}
\Box =-\partial ^A\partial _A \\
\\
\bigtriangleup _1=\lambda _a\overline{\lambda }_b\partial ^b\overline{%
\partial }^a\quad ,\quad \bigtriangleup _2=\overline{\lambda }_a\lambda
_b\partial ^{ab}\partial _{cd}\overline{\partial }^c\partial ^d
\end{array}
\end{equation}
where $\partial _{ab}=\left( \sigma ^A\right) _{ab}\partial _A$. Casimir
operators of the Poincar\'e group in representation (\ref{i}) can be
expressed through the Laplace operators as follows
\begin{equation}
\label{m}
\begin{array}{c}
C_1=
{\bf P}^A{\bf P}_A=-\Box \\  \\
C_2=\frac 1{24}{\bf W}^{ABC}{\bf W}_{ABC}=\bigtriangleup _2+\Box
\bigtriangleup _1\quad ,\quad C_3=\frac 1{64}{\bf W}^A{\bf W}_A={\bf %
\bigtriangleup }_1\bigtriangleup _2+2\Box \bigtriangleup _1
\end{array}
\end{equation}
where ${\bf W}^A=\epsilon ^{ABCDEF}{\bf M}_{BC}{\bf M}_{DE}{\bf P}_F$ , $%
{\bf W}^{ABC}=\epsilon ^{ABCDEF}{\bf M}_{DE}{\bf P}_F$ are Pauli-Lubanski
vector and tensor respectively.

In what follows we present another covariant parametrization of $M^{12}$ in
terms of a non-zero light-like vector $b^A$ and anti-self-dual tensor $%
h^{ABC}$ constrained by the relations
\begin{equation}
\label{n}
\begin{array}{c}
b^Ab_A=0\quad ,\quad b^A\sim ab^A\quad ,\quad h^{ABC}\sim ah^{ABC}\quad
,\quad a\in R\backslash \left\{ 0\right\} \\
\\
h^{ABC}=-\frac 16\epsilon ^{ABCDEF}h_{DEF}\qquad ,\qquad b_Ah^{ABC}=0 \\
\\
h^{ABC}h_{CDE}=\frac 14\delta ^{[A}{}_{[D}b^{B]}b_{E]}
\end{array}
\end{equation}
(Here the anti-self-dual tensor $h^{ABC}$ is chosen real that is
always possible in $R^{5,1}$.) As a matter of fact, the first two relations
imposed on $b^A$ define $S^4$ as a projective light-cone. With the use of
Lorentz transformations each point on $S^4$ can be brought into another one
parametrized by the vector $\stackrel{o}{b}\!^A=\left( 1,0,0,0,0,1\right) $.
By substituting $\stackrel{o}{b}\,^A$ into the fourth equation one reduces
the ten components of $h^{ABC}$ to the three independent values, for
instance $h_{012},h_{013},h_{014}$. Then the last equation takes the form
\begin{equation}
\left( h_{012}\right) ^2+\left( h_{013}\right) ^2+\left( h_{014}\right)
^2=\frac 14
\end{equation}
i.e. it defines the two-sphere $S^2$. In such a manner we recover the local
structure of $CP^3$ discussed above (\ref{f}). The relationship between
these two parametrizations may be established explicitly with the use of the
following Fierz identity:
\begin{equation}
\label{o}\overline{\lambda }_a\lambda _b=\frac 14\overline{\lambda }%
\widetilde{\sigma }_A\lambda \left( \sigma ^A\right) _{ab}+\frac 1{12}%
\overline{\lambda }\widetilde{\sigma }_{ABC}\lambda \left( \sigma
^{ABC}\right) _{ab}
\end{equation}

Defining $b^A$ and $h^{ABC}$ through $\overline{\lambda }_a$,$\lambda _a$ as
\begin{equation}
\label{p}b^A=\overline{\lambda }\widetilde{\sigma }{}^A\lambda \qquad
,\qquad h^{ABC}=i{}\overline{\lambda }\widetilde{\sigma }{}^{ABC}\lambda
\end{equation}
one can get (\ref{n}). The Poincar\'e generators (\ref{i}) and Laplace
operators (\ref{m}) can straightforwardly be rewritten in terms of $b^A$ and
$h^{ABC}$ but we omit the explicit expressions here since in what follows
the spinor parametrization of $CP^3$ will be mainly used.

\section{Action functional in the first order formalism and classical
dynamics}

We proceed to the derivation an action functional governing the point
particle dynamics on $M^{12}.$ The main dynamical principle underlying our
construction is the requirement of identical (off-shell) conservation for
the classical counterparts of three Casimir operators (\ref{m}).

As a starting point, consider the phase space $T^{*}(R^{5,1}\times C^4)$
parametrized by the coordinates $x^A,\lambda _a,\overline{\lambda }_b$ and
their conjugated momenta $p_A,\pi ^a,\overline{\pi }^b$ satisfying the
canonical Poisson-bracket relations
\begin{equation}
\label{q}\left\{ x^A,p_B\right\} =\delta ^A\!_B\quad ,\quad \left\{ \lambda
_a,\pi ^b\right\} =\delta ^b\!_a\quad ,\quad \left\{ \overline{\lambda }_a,%
\overline{\pi }^b\right\} =\delta ^b\!_a
\end{equation}
Obviously, the action of the Poincar\'e group on $M^{12}$ (\ref{i}) is lifted up
to the canonical action on $T^{*}(R^{5,1}\times C^4).$ This action induces a
special representation of the Poincar\'e group on the space of smooth
functions $F$ over the phase space, and the corresponding infinitesimal
transformations can be written via the Poisson brackets as follows
\begin{equation}
\label{r}\delta F=\left\{ F,-a^AP_A+\frac 12K^{AB}J_{AB}\right\}
\end{equation}
Here $a^A$ and $K^{AB}=-K^{BA}$ are the parameters of translations and
Lorentz transformations, respectively, and the Hamilton generators read
\begin{equation}
\label{s}P_A=p_A\qquad ,\qquad J_{AB}=x_Ap_B-x_Bp_A+M_{AB}
\end{equation}
where the spinning part of Lorentz generators is given by%
$$
M_{AB}=-\pi \sigma _{AB}\lambda +\ c.c.
$$
The phase-space counterparts of Casimir operators associated with the
generators (\ref{s}) can be readily obtained from (\ref{m}) by making formal
replacements: $\partial _A\rightarrow p_A,\partial ^a\rightarrow \pi ^a,%
\overline{\partial }^a\rightarrow \overline{\pi }^a$. The result is
\begin{equation}
\label{t}
\begin{array}{c}
C_1=p^2 \\
\\
C_2=p^2\left( \overline{\pi }\lambda \right) \left( \pi \overline{\lambda }%
\right) -\left( \overline{\pi }p\pi \right) \left( \overline{\lambda }%
p\lambda \right) \quad ,\quad C_3=\left( \overline{\pi }\lambda \right)
\left( \pi \overline{\lambda }\right) \left( \overline{\pi }p\pi \right)
\left( \overline{\lambda }p\lambda \right)
\end{array}
\end{equation}
As is seen the Casimir functions $C_2,C_3$ are unambiguously expressed via
the classical analogs of Laplace operators (\ref{l})
\begin{equation}
\label{u}\bigtriangleup _1=\left( \overline{\pi }\lambda \right) \left( \pi
\overline{\lambda }\right) \quad ,\quad \bigtriangleup _2=\left( \overline{%
\pi }p\pi \right) \left( \overline{\lambda }p\lambda \right)
\end{equation}
and, thereby, one may require the identical conservation of $\bigtriangleup
_1,\bigtriangleup _2$ instead of $C_2,C_3$.

Let us now introduce the set of five first-class constraints, three of which
are dynamical
\begin{equation}
\label{v}
\begin{array}{c}
T_1=p^2+m^2\approx 0 \\
\\
T_2=\bigtriangleup _1+\delta _1^2\approx 0\qquad ,\qquad T_3=\bigtriangleup
_2+m^2\delta _2^2
\end{array}
\end{equation}
and the other are kinematical
\begin{equation}
\label{w}T_4=\pi \lambda \approx 0\qquad ,\qquad T_5=\overline{\pi }%
\overline{\lambda }\approx 0
\end{equation}
Here parameter $m$ is identified with the mass of the particle, while the
parameters $\delta _1,\delta _2$ relate to the particle's spin. The role of
kinematical constraints is to make the Hamiltonian reduction of the extended
phase space to the cotangent bundle $T^{*}\left( M^{12}\right) $. In
configuration space these constraints generate the equivalence relation (\ref
{h}) with respect to the Poisson brackets (\ref{q}). The constraints
$T_1,T_2,T_3$ determine the dynamical content of the model and lead to the
unique choice for action functional.

>From (\ref{v}) it follows that on the constraint surface the conserved
charges $\bigtriangleup _1$ and $\bigtriangleup _2$ are limited to be
negative (or zero) constants. These restrictions are readily seen from the
following simple reasons. Let us introduce the set of three $p$-transversal
tensors
\begin{equation}
\label{x}
\begin{array}{c}
W_{ABC}=\epsilon _{ABCDEF}J^{DE}p^F\ \ ,\ \ W_A=\epsilon
_{ABCDEF}J^{BC}J^{DE}p^F\ \ ,\ \  \\
\\
V_A=M_{AB}p^B
\end{array}
\end{equation}
Since the $p$ is a time-like vector (\ref{v}) the full contraction of each
introduced tensor with itself should be non-negative. Then one may check
that the following relations take place
\begin{equation}
\label{y}
\begin{array}{c}
W_{ABC}W^{ABC}=p^2\bigtriangleup _1-\bigtriangleup _2\geq 0\quad ,\quad
W_AW^A=\bigtriangleup _1\bigtriangleup _2\geq 0, \\
\\
V_AV^A=-p^2\bigtriangleup _1-\bigtriangleup _2\geq 0
\end{array}
\end{equation}

Resolving these inequalities we come to the final relation:
\begin{equation}
\label{z}\bigtriangleup _2\leq m^2\bigtriangleup _1\leq 0
\end{equation}
which in turn implies that $\left| \delta _2\right| \geq \left| \delta
_1\right| $. Thus, the set of constraints (\ref{v}) leads to the
self-consistent classical dynamics only provided that the rel.(\ref{z}) holds
true.

Assuming the theory to be reparametrization invariant, the Hamiltonian of
the model is a linear combination of the constraints and the first-order
(Hamilton) action takes the form:
\begin{equation}
\label{aa}S_H=\int d\tau \left\{ p_A\stackrel{.}{x}^A+\pi ^a\stackrel{.}{%
\lambda }_a+\overline{\pi }^a\stackrel{.}{\overline{\lambda }}%
_a-\sum\limits_{i=1}^5e_iT^i\right\}
\end{equation}
Here $\tau $ is the evolution parameter, $e_i$ are the Lagrange multipliers
associated to the constraints with $e_4=\overline{e}_5$. Varying (\ref{aa})
one gets the following equations of motion:
\begin{equation}
\label{ab}
\begin{array}{c}
\dot \lambda _a=e_2\left(
\overline{\pi }\lambda \right) \overline{\lambda }_a+e_3\left( \overline{%
\lambda }p\lambda \right) \overline{\pi }^bp_{ba}+e_4\lambda _a \\  \\
\dot \pi ^a=-e_2\left( \pi
\overline{\lambda }\right) \overline{\pi }^a-e_3\left( \overline{\pi }p\pi
\right) \overline{\lambda }_bp^{ba}-e_4\pi ^a \\  \\
\dot x^A=2e_1p^A+e_3\left\{ \left(
\overline{\pi }\sigma ^A\pi \right) \left( \overline{\lambda }p\lambda
\right) +\left( \overline{\pi }p\pi \right) \left( \overline{\lambda }%
\widetilde{\sigma }^A\lambda \right) \right\} \\  \\
\dot p_A=0
\end{array}
\end{equation}
Despite the quite nonlinear structure, the equations are found to be
completely integrable with arbitrary Lagrange multipliers. This fact is not
surprising as the model, by construction, describes a free relativistic
particle possessing a sufficient number of symmetries.

In the spinning sector the corresponding solution looks like:
$$
\begin{array}{c}
\displaystyle{\ \lambda _a=e}^{E_4}{\cos \left( m^2E_3\delta _2\right)
\left( \cos \left( E_2\delta _1\right) \lambda _a^0+\frac{\sin \left(
E_2\delta _1\right) }{\delta _1}\left( \overline{\pi }_0\lambda ^0\right)
\overline{\lambda }_a^0\right) +} \\  \\
\displaystyle{\ +e}^{E_4}{\frac{\left( \overline{\lambda }^0p\lambda
^0\right) }{m^2}\frac{\sin \left( m^2E_3\delta _2\right) }{\delta _2}%
p_{ab}\left( \frac{\sin \left( E_2\delta _1\right) }{\delta _1}\left(
\overline{\pi }_0\lambda ^0\right) \pi _0^b-\cos \left( E_2\delta _1\right)
\overline{\pi }_0^b\right) }
\end{array}
$$
\begin{equation}
\label{ac}{}
\end{equation}
$$
\begin{array}{c}
\displaystyle{\ \pi ^a=e}^{-E_4}{\cos \left( m^2E_3\delta _2\right) \left(
\cos \left( E_2\delta _1\right) \pi _0^a-\frac{\sin \left( E_2\delta
_1\right) }{\delta _1}\left( \pi _0\overline{\lambda }^0\right) \overline{%
\pi }_0^a\right) +} \\  \\
\displaystyle{\ +e}^{-E_4}{\frac{\left( \overline{\pi }_0p\pi _0\right) }{m^2%
}\frac{\sin \left( m^2E_3\delta _2\right) }{\delta _2}p^{ab}\left( \frac{%
\sin \left( E_2\delta _1\right) }{\delta _1}\left( \pi _0\overline{\lambda }%
^0\right) \lambda _b^0+\cos \left( E_2\delta _1\right) \overline{\lambda }%
_b^0\right) }
\end{array}
$$
and for the space-time evolution one gets
\begin{equation}
\label{ad}
\begin{array}{c}
p^A=p_0^A \\
\\
x^A\left( \tau \right) =x_0^A+2\left( E_1+E_3\delta _2^2\right)
p_0^A-m^{-2}V^A\left( \tau \right) \\
\\
V^A\left( \tau \right) =V_1^A\cos \left( 2m^2E_3\delta _2\right) +V_2^A\sin
\left( 2m^2E_3\delta _2\right)
\end{array}
\end{equation}
Here $E_i\left( \tau \right) =\int\limits_0^\tau d\tau e_i(\tau )$, vector $%
V^A$ is defined as in (\ref{x}) and the initial data $\lambda _a^0=\lambda
_a\left( 0\right) \ ,\ \pi _0^a=\pi ^a\left( 0\right) \ ,\ p_0^A$ are
assumed to be chosen on the surface of constraints (\ref{v}), (\ref{w}).

Let us briefly discuss the obtained solution. First of all, one may resolve
the kinematical constraints (\ref{w}) by imposing the gauge fixing
conditions of the form
\begin{equation}
\label{ae}e_4=e_5=0\qquad ,\qquad \lambda _0=1 \, ,
\end{equation}
so that $\lambda _i,i=1,2,3$ can be treated as the local coordinates on $%
CP^3 $. Then from (\ref{ac}), (\ref{ad}) we see that the motion of the point
particle on $M^{12}$ is completely determined by an independent evolution of
the three Lagrange multipliers $e_1,e_2,e_3$. The presence of two additional
gauge invariances in comparison with spinless particle case causes the
conventional notion of particle world line, as the geometrical set of
points, to fail. Instead, one has to consider the class of gauge equivalent
trajectories on $M^{12}$ which, in the case under consideration, is identified
with three-dimensional surface, parametrized by $e_1,e_2,e_3$ . The
space-time projection of this surface is represented by the two-dimensional
tube of radius $\rho =\sqrt{\delta _2^2-\delta _1^2}$ along the particle's
momenta $p$ as is seen from the explicit expression (\ref{ad}). This fact
becomes more clear in the rest reference system $\stackrel{\circ }{p}_A=(m,%
\stackrel{\rightarrow }{0})$ after identifying of the evolution parameter $%
\tau $ with the physical time by the law
\begin{equation}
\label{af}x^0=c\tau
\end{equation}
Then eq. (\ref{ad}) reduces to
\begin{equation}
\label{ag}\stackrel{\rightarrow }{x}(\tau )=m^{-2}\stackrel{\rightarrow }{V}%
\left( \tau \right) =\stackrel{\rightarrow }{V}_1\cos \left( 2m^2E_3\delta
_2\right) +\stackrel{\rightarrow }{V}_2\sin \left( 2m^2E_3\delta _2\right)
\end{equation}
where, in accordance with (\ref{y}) $\stackrel{\rightarrow }{V}%
{}^2=m^2\left( \delta _2^2-\delta _1^2\right) $ and hence
\begin{equation}
\label{ah}\stackrel{\rightarrow }{V}_1\!^2=\stackrel{\rightarrow }{V}%
_2{}^2=\delta _2^2-\delta _1^2\ ,\ (\stackrel{\rightarrow }{V}_1,\stackrel{%
\rightarrow }{V}_2)=0
\end{equation}
The rest gauge arbitrariness, related to the Lagrange multiplier $e_3$, causes
that, in each moment of time, the space-time projection of the motion is
represented by a circle of radius $\rho $. This means that after accounting
spin, the relativistic particle ceases to be localized in a certain point of
Minkowski space but represents a string-like configuration contracting to
the point only provided that $\delta _1=\delta _2$.

Finally, let us discuss the structure of the physical observables of the
theory. Each physical observable $A$ being a gauge-invariant function on the
phase space should meet the requirements:
\begin{equation}
\label{ai}\left\{ A,T_i\right\} =0\qquad ,\qquad i=1,..,5
\end{equation}
Due to the obvious Poincar\'e invariance of the constraint surface, the
generators (\ref{s}) automatically satisfy (\ref{ai}) and thereby they are
the observables. On the other hand, it is easy to compute that the dimension
of the physical phase space of the theory is equal to 18. Thus the physical
subspace may covariantly be parametrized by 21 Poincar\'e generator subject
to 3 conditions (\ref{t}), and as a result, any physical observable proves to
be a function of the generators (\ref{s}) modulo constraints. So a general
solution of (\ref{ai}) reads
\begin{equation}
\label{aj}A=A\left( J_{AB},P_C\right) +\sum\limits_{i=1}^5\alpha _iT_i
\end{equation}
$\alpha _i$, being an arbitrary function of phase space variables.

In fact, this implies that the physical phase space of the model is embedded
in the linear space of the Poincar\'e algebra through the constraints (\ref
{v}) and therefore coincides with some coadjoint orbit ${\cal O}$ of the
Poincar\'e group.

\section{Second-order formalism}

In order to obtain a second-order formulation for the model one may proceed
in the standard manner by eliminating the momenta $p_A,\pi ^a,\overline{\pi }%
^a$ and the Lagrange multipliers $e_i$ from the Hamiltonian action (\ref{aa}%
) resolving equations of motion:
\begin{equation}
\label{ak}\frac{\delta S}{\delta p_A}=\frac{\delta S}{\delta \pi ^a}=\frac{%
\delta S}{\delta \overline{\pi }^a}=\frac{\delta S}{\delta e_i}=0
\end{equation}
with respect to the momenta and the multipliers. The corresponding
Lagrangian action will be invariant under global Poincar\'e transformation
and will possess five gauge symmetries associated with first-class
constraints (\ref{v}). The presence of kinematical ones will result in the
invariance of Lagrangian action under the local $\lambda $-rescalings: $%
\lambda _a\rightarrow \alpha \lambda _a$. At the same time, by construction,
among the gauge transformations related to the dynamical constraints will
necessarily be the one corresponding to reparametrizations of the particle
world-line $\tau \rightarrow \tau ^{\prime }\left( \tau \right) $.

It turns out, however, that the straightforward resolution of eqs. (\ref
{ak}) is rather cumbersome. Fortunately, in the case in hand there is
another way to recover the covariant second-order formulation exploiting the
symmetry properties of the model. Namely, we can start with the most general
Poincar\'e and reparametrization invariant ansatz for the Lagrange action
and specify it, by requiring the model to be equivalent to that
described by the constraints (\ref{v}).

As a first step we classify all the Poincar\'e invariants of the
world-line being functions over the tangent bundle $TM^{12}$. One may easily
verify that there are only three expressions possessing these properties
\begin{equation}
\label{al}\stackrel{.}{x}^2\qquad ,\qquad \xi =\frac{(\stackrel{.}{\lambda }%
\stackrel{.}{x}\lambda )(\stackrel{.}{\overline{\lambda }}\stackrel{.}{x}%
\overline{\lambda })}{\stackrel{.}{x}^2\left( \overline{\lambda }\stackrel{.%
}{x}\lambda \right) ^2}\qquad ,\qquad \eta =\frac{\epsilon ^{abcd}\stackrel{.%
}{\lambda }_a\overline{\lambda }_b\stackrel{.}{\overline{\lambda }}_c\lambda
_d}{\left( \overline{\lambda }\stackrel{.}{x}\lambda \right) ^2}
\end{equation}
Notice that $\xi $ and $\eta $ are invariant under reparametrizations as
well as under the local $\lambda $-rescalings (\ref{h}), so the kinematical
constraints (\ref{w}) are automatically accounted

Then the most general Poincar\'e and reparametrization invariant Lagrangian
on $M^{12}$ reads
\begin{equation}
\label{am}{\cal L=}\sqrt{-\stackrel{.}{x}^2F\left( \xi ,\eta \right) }
\end{equation}
where $F$ is an arbitrary function.

The particular form of the function $F$ entering (\ref{am}) may be found
from the requirement that the Lagrangian is to lead to the Hamilton
constraints (\ref{v}). The substitution of the canonical momenta
\begin{equation}
\label{an}p_A=\frac{\partial {\cal L}}{\partial \stackrel{.}{x}^A}\qquad
,\qquad \pi ^a=\frac{\partial {\cal L}}{\partial \stackrel{.}{\lambda }_a}%
\qquad ,\qquad \overline{\pi }^a=\frac{\partial {\cal L}}{\partial \stackrel{%
.}{\overline{\lambda }}_a}
\end{equation}
to the dynamical constraints $T_1$ and $T_2$ gives the following equations
\begin{equation}
\label{ao}
\begin{array}{c}
\displaystyle{\frac{\partial {\cal L}}{\partial \stackrel{.}{x}^A}\frac{%
\partial {\cal L}}{\partial \stackrel{.}{x}_A}+m^2=0\Rightarrow } \\  \\
\displaystyle{\Rightarrow F^2+\xi \left( \xi +\eta \right) \left( \frac{%
\partial F}{\partial \xi }\right) ^2-2\xi \frac{\partial F}{\partial \xi }%
-2\eta \frac{\partial F}{\partial \eta }+2\xi \eta \frac{\partial F}{%
\partial \xi }\frac{\partial F}{\partial \eta }-m^2F=0}
\end{array}
\end{equation}
and
\begin{equation}
\label{ap}\frac{\partial {\cal L}}{\partial \stackrel{.}{\overline{\lambda }}%
_a}\lambda _a\frac{\partial {\cal L}}{\partial \stackrel{.}{\lambda }_b}%
\overline{\lambda }_b+\delta _1^2=0\Rightarrow \left( \frac{\partial F}{%
\partial \xi }\right) ^2+\delta _1^2F=0
\end{equation}
The integration of these equations results with
\begin{equation}
\label{aq}F=\left( 2\delta _1\sqrt{-\xi }+\sqrt{m^2-4\delta _1^2\eta +4A%
\sqrt{\eta }}\right) ^2\quad ,
\end{equation}
$A$ being arbitrary constant of integration. The account of the rest
constraint $T_3$ does not contradict the previous equations, but
determines the value of $A$ as
\begin{equation}
\label{ar}A=m\sqrt{\delta _2^2-\delta _1^2}
\end{equation}
Putting altogether, we come with the Lagrangian
$$
\displaystyle{{\cal L}=\sqrt{-\stackrel{.}{x}^2\left( m^2-4\delta _1^2\frac{%
\epsilon ^{abcd}\stackrel{.}{\lambda }_a\overline{\lambda }_b\stackrel{.}{%
\overline{\lambda }}_c\lambda _d}{\left( \overline{\lambda }\stackrel{.}{x}%
\lambda \right) ^2}+4m\sqrt{\left( \delta _2^2-\delta _1^2\right) \frac{%
\epsilon ^{abcd}\stackrel{.}{\lambda }_a\overline{\lambda }_b\stackrel{.}{%
\overline{\lambda }}_c\lambda _d}{\left( \overline{\lambda }\stackrel{.}{x}%
\lambda \right) ^2}}\right) }}+
$$
\begin{equation}
\label{as}{}
\end{equation}
$$
\displaystyle{+2\delta _1\left| \frac{\stackrel{.}{\lambda }\stackrel{.}{x}%
\lambda }{\overline{\lambda }\stackrel{.}{x}\lambda }\right| }
$$
It should be stressed that the parameters $\delta _1$ and $\delta _2$
entering the Lagrangian are dimensional and cannot be made dimensionless by
redefinitions involving only the mass of the particle and the speed of light
$c$. Whereas, using the Planck constant we may set
\begin{equation}
\label{at}\delta _1=\frac \hbar c\kappa _1\qquad ,\qquad \delta _2=\frac
\hbar c\kappa _2
\end{equation}
where $\kappa _1$ and $\kappa _2$ are already arbitrary real numbers
satisfying the inequality $\left| \kappa _1\right| \leq \left| \kappa
_2\right| $. Turning back to the question of particle motion (see (\ref{ad})
and below) we also conclude that the radius $\rho $ of the tube,
representing the particle propagation in Minkowski space is proportional to $%
\hbar $. So, this ''non-local'' behavior of the particle is caused by spin
which manifests itself as a pure quantum effect disappearing in the
classical limit $\hbar \rightarrow 0$.

As is seen, for a given non-zero, spin the Lagrangian (\ref{as}) has a
complicated structure involving radicals and, hence, the reality condition
for ${\cal L}$ requires special consideration. Similar to the spinless case,
the space-time causality implies that
\begin{equation}
\label{au}\stackrel{.}{x}^2<0\qquad ,\qquad \stackrel{.}{x}^0>0
\end{equation}
Then expression (\ref{as}) is obviously well-defined only provided that
\begin{equation}
\label{av}
\begin{array}{c}
\eta \geq 0 \\
\\
m^2-4\delta _1^2\eta +4m\sqrt{\left( \delta _2^2-\delta _1^2\right) \eta }%
\geq 0
\end{array}
\end{equation}
As will be seen below the first inequality is always fulfilled, while the
second condition is equivalent to
\begin{equation}
\label{aw}0\leq \eta \leq \frac{m^2}{4\delta _1^4}\left( \delta _2+\sqrt{%
\delta _2^2-\delta _1^2}\right) ^2
\end{equation}
Together, eqs. (\ref{au}), (\ref{aw}) may be understood as the full set of
causality conditions for the model of massive spinning particle.

Passing to the vector parametrization of the configuration space in terms of
$b_A$ and $h_{ABC}$ the basic invariants $\eta $ and $\xi $ take the form
\begin{equation}
\label{ax}
\begin{array}{c}
\displaystyle{\xi =-\frac{4\stackrel{.}{x}_A\stackrel{.}{h}^{ABC}\stackrel{.%
}{h}_{BCD}\stackrel{.}{x}^D+\stackrel{.}{x}^2\stackrel{.}{b}^2-4\left(
\stackrel{.}{x}\stackrel{.}{b}\right) ^2}{16\stackrel{.}{x}^2\left(
\stackrel{.}{x}b\right) ^2}} \\  \\
\displaystyle{\eta =\frac{\stackrel{.}{b}^2}{4\left( \stackrel{.}{x}b\right)
^2}}
\end{array}
\end{equation}
and the corresponding Lagrangian reads
\begin{equation}
\label{ay}
\begin{array}{c}
\displaystyle{{\cal L}=\sqrt{-\stackrel{.}{x}^2\left( m^2-\delta _1^2\frac{%
\stackrel{.}{b}^2}{\left( \stackrel{.}{x}b\right) ^2}+2m\sqrt{\left( \delta
_2^2-\delta _1^2\right) \frac{\stackrel{.}{b}^2}{\left( \stackrel{.}{x}%
b\right) ^2}}\right) }}+ \\  \\
\displaystyle{+\delta _1\sqrt{\frac{4\stackrel{.}{x}_A\stackrel{.}{h}^{ABC}%
\stackrel{.}{h}_{BCD}\stackrel{.}{x}^D+\stackrel{.}{x}^2\stackrel{.}{b}%
^2-4\left( \stackrel{.}{x}\stackrel{.}{b}\right) ^2}{4\left( \stackrel{.}{x}%
b\right) ^2}}}
\end{array}
\end{equation}
where the holonomic constraints (\ref{n}) are assumed to hold. In view of (%
\ref{ax}) the condition (\ref{av}) becomes evident since $\stackrel{.}{b}^A$
is orthogonal to the light-like vector $b^A$ and thereby is space-(or
light-) like. Recalling that the vector $b^A$ parametrizes $S^4$,
condition (\ref{aw}) forbids the particle to move with arbitrary large
velocity not only in Minkowski space but also on the sphere $S^4$.

Classically the parameters $\delta _1$ and $\delta _2$ can be chosen to be
arbitrary numbers subject only to the restriction $\left| \delta _1\right|
\leq \left| \delta _2\right| $. There are, however, two special cases: $%
\delta _1=\delta _2=0$ and $\delta _1=0$ when the Lagrangian (\ref{ay}) is
considerably simplified. The former option is of no interest as it
corresponds
to the case of spinless massive particle, while the latter leads to the
following Lagrangian
\begin{equation}
\label{az}{\cal L}=\sqrt{-\stackrel{.}{x}^2\left( m^2+2m\delta _2\sqrt{\frac{%
\stackrel{.}{b}{}^2}{\left( \stackrel{.}{x}b\right) ^2}}\right) }
\end{equation}
which is the direct six-dimensional generalization of the $\left( m,s\right)
$-particle model proposed earlier \cite{c13} for $D=4$. The configuration
space of the model (\ref{az}) is represented by the direct product of
Minkowski space $R^{5,1}$ and four-dimensional sphere $_{}S^4$ parametrized
by the light-like vector $b^A$. It is easy to see that the reduced model
cannot describe arbitrary spins, since the third Casimir operator (\ref{m}),
being constructed from the Poincar\'e generators acting on $R^{5,1}\times
S^4 $, vanishes identically. As will be seen below the quantization of this
case leads to the irreducible representations of the Poincar\'e group
realized on totally symmetric tensor fields on Minkowski space.

\section{Generalization to the curved background}

So far we discussed the model of spinning particle living on the flat
space-time. In this section, we will try to generalize it to the case of
curved background. For these ends one can replace the configuration space $%
M^{12}$ by ${\cal M}^6\times CP^3$ where ${\cal M}^6$ is a curved
space-time. Now the action functional should be generalized to remain
invariant under both general coordinate transformations on ${\cal M}^6$ and
local Lorentz transformations on $CP^3$. Let $e_m{}^A$ and $\omega _{mAB}$
be the vielbein and the torsion free spin connection respectively. The
minimal covariantization of the Lagrangian (\ref{as}) gives
$$
\displaystyle{\!{\cal L}=\!\sqrt{-\stackrel{.}{x}^2\!\left( \!m^2-\!4\delta
_1^2\frac{\epsilon ^{abcd}\stackrel{\bullet }{\lambda }_a\overline{\lambda }%
_b\stackrel{\bullet }{\overline{\lambda }}_c\lambda _d}{\left( \stackrel{.}{x%
}^me_m{}^A\left( \overline{\lambda }\sigma _A\lambda \right) \right) ^2}+\!4m%
\sqrt{\left( \delta _2^2-\delta _1^2\right) \frac{\epsilon ^{abcd}\stackrel{%
\bullet }{\lambda }_a\overline{\lambda }_b\stackrel{\bullet }{\overline{%
\lambda }}_c\lambda _d}{\left( \stackrel{.}{x}^me_m{}^A\left( \overline{%
\lambda }\sigma _A\lambda \right) \right) ^2}}\right) }}+
$$
\begin{equation}
\label{ca}{}
\end{equation}
$$
\displaystyle{+2\delta _1\left| \frac{\stackrel{.}{x}^me_m{}^A\left(
\stackrel{\bullet }{\lambda }\sigma _A\lambda \right) }{\stackrel{.}{x}%
^me_m{}^A\left( \overline{\lambda }\sigma _A\lambda \right) }\right| }
$$
where
\begin{equation}
\label{cb}\stackrel{\bullet }{\lambda }_a=\stackrel{.}{\lambda }_a-\frac 12%
\stackrel{.}{x}^m\omega _{mAB}(\sigma ^{AB})_a{}^b\lambda _b
\end{equation}
is the Lorentz covariant derivative along the particle's world line.

Proceeding to the Hamilton formalism one gets the set of five constraints $%
T_i^{^{\prime }}$, $i=1...5$ which may be obtained from $T_i$ (\ref{v},\ref
{w}) by replacing $p_A\rightarrow \Pi _A$, where
\begin{equation}
\label{cc}\Pi _A=e_A{}^m\left( p_m+\frac 12\omega _{mCD}M^{CD}\right)
\end{equation}
Here $e_A{}^m$ is the inverse vielbein and $M^{CD}$ is the spinning part of
Lorentz generators (\ref{s}). The generalized momentum $\Pi _A$ satisfies
the following Poisson brackets relation:
\begin{equation}
\label{cd}\left\{ \Pi _A,\Pi _B\right\} =\frac 12R_{ABCD}M^{CD}
\end{equation}
$R_{ABCD}$ being the curvature tensor of ${\cal M}^6$. Now it is easy to
find that
\begin{equation}
\label{ce}
\begin{array}{c}
\left\{ T_1^{^{\prime }},T_3^{^{\prime }}\right\} =R_{ABCD}q^A\Pi ^BM^{CD}
\\
\\
q^A=(\overline{\lambda }\sigma ^A\lambda )(\overline{\pi }\Pi \pi )+(%
\overline{\lambda }\Pi \lambda )(\overline{\pi }\sigma ^A\pi )
\end{array}
\end{equation}
The other Poisson brackets of the constraints are equal to zero. So, in
general, the constraints $T_1^{^{\prime }},T_3^{^{\prime }}$ are of the
second class, which implies that switching on an interaction destroys the
first class constraints algebra and, hence, gives rise to unphysical degrees
of freedom in the theory. What is more, the Lagrangian (\ref{ca}) is
explicitly invariant under reparametrizations of the particle's world line,
while the gauge transformations, associated with the remaining first class
constraints $T_2^{^{\prime }},T_4^{^{\prime }},T_5^{^{\prime }}$, do not
generate the full reparametrizations of the theory (the space-time
coordinates $x^m$ on ${\cal M}^6$ remain intact ). The last fact indicates
that the equations of motion derived from (\ref{ca}) are contradictory. Thus
the interaction with external gravitational field is self-consistent only
provided that r.h.s. of (\ref{ce}) vanishes. This requirement leads to some
restrictions on curvature tensor. Namely, with the use of the identity $%
M^{AB}q_B\approx 0$ one may find that (\ref{ce}) is equal to zero if and
only if $R_{ABCD}$ has the form
\begin{equation}
\label{cf}R_{ABCD}=\frac R{30}\left( \eta _{AC}\eta _{BD}-\eta _{AD}\eta
_{BC}\right)
\end{equation}
where $R$ is a constant (the scalar curvature of the manifold ${\cal M}^6$).
So the minimal coupling to gravity is self-consistent only provided that $%
{\cal M}^6$ is the space of constant curvature.

Concluding this section let us also remark that the Lagrangian (\ref{ca})
may be obtained using the group theoretical principles outlined in the
introduction. To this end one should replace the Poincar\'e group by $%
SO\left( 5,2\right) $ or $SO\left( 6,1\right) $ depending on $R<0$ or $R>0$.
(Cf. see \cite{c13})

\section{Quantization}

In Sect. 3 we have seen that the model is completely characterized, at the
classical level, by the algebra of observables associated with the phase
space generators of the Poincar\'e group. We have shown that the observables
${\cal A}=(P_A,J_{AB})$ constitute the basis, so that any gauge invariant
value of the theory can be expressed via the elements of ${\cal A}$.

To quantize this classical system means to construct an irreducible unitary
representation
\begin{equation}
\label{ba}r:{\cal A\rightarrow }End\ {\cal H}
\end{equation}
of the Lie algebra ${\cal A}$ in the algebra $End\ {\cal H}$ of linear
self-adjoint operators in a Hilbert space where the physical subspace ${\cal %
H}$ is identified with the kernel of the first-class constraint operators.
Here by a Lie algebra representation $r$ we mean a linear mapping from $%
{\cal A}$ into $End\ {\cal H}$ such that
\begin{equation}
\label{bb}r(\{{f,g\}})=-i[r(f),r(g)]\qquad ,\qquad \forall \ f,g\in {\cal A}
\end{equation}
where $[r(f),r(g)]$ is the usual commutator of Hermitian operators $r(f)$, $%
r(g).$ Unitarity means that the canonical transformations of the model's
phase space generated by observables from ${\cal A}$ should correspond to
unitary transformations on ${\cal H.}$ Besides that we should supply the
algebra ${\cal A}$ by the central element 1 and normalize $r$ by the
condition
\begin{equation}
\label{bc}r(1)=id
\end{equation}
i.e. the constant function equal to 1 corresponds under $r$ to the identity
operator on ${\cal H}$.

Now{\it \ }it is seen{\it \ that\ the quantization of the model is reduced
to the construction of the unitary irreducible representation of the
Poincar\'e group with the given quantum numbers fixed by the constraints }(%
\ref{v}, \ref{w}).

Within the framework of the covariant operatorial quantization the Hilbert
space of physical states ${\cal H}$ is embedded into the space of smooth
scalar functions on $R^{5,1}\times C^4$ and the phase space variables $%
x^A,p_A,\lambda _a,\pi ^a$ are considered to be Hermitian operators subject
to the canonical commutation relations.

In the ordinary coordinate representation
\begin{equation}
\label{bd}p_A\rightarrow -i\partial _A\qquad ,\qquad \pi ^a\rightarrow
-i\partial ^a\qquad ,\qquad \overline{\pi }^a\rightarrow -i\overline{%
\partial }^a
\end{equation}
the Hermitian generators of the Poincar\'e group (observables) take the form
\begin{equation}
\label{be}{\bf P}_A=-i\partial _A\quad ,\qquad {\bf M}_{AB}=-i\left(
x_A\partial _B-x_B\partial _A+\left( \sigma _{AB}\right) _a\!^b\left(
\lambda _b\partial ^a+\overline{\lambda }_b\overline{\partial }^a\right)
\right)
\end{equation}
By contrast, the quantization of the first-class constraints is not so
unambiguous. As is seen from the explicit expressions (\ref{v}, \ref{w})
there is an inherent ambiguity in the ordering of operators $\widehat{%
\lambda }_a,\widehat{\pi }^b$ and $\widehat{\overline{\lambda }}_a,\widehat{%
\overline{\pi }}^b$. Luckily as one may verify, the different ordering
prescription for the above operators results only in renormalization of the
parameters $\delta _1^2,\delta _2^2$ and modification of the kinematical
constraints by some constants $n$ and $m$. Thus, in general, (after omitting
inessential multipliers) the quantum operators for the first-class
constraints may be written as
\begin{equation}
\label{bf}
\begin{array}{c}
\widehat{T}_1=\Box -m^2\quad ,\quad \widehat{T}_2=\bigtriangleup _1-\delta
_1^{^{\prime }2}\quad ,\quad \widehat{T}_3=\bigtriangleup _2-\delta
_2^{^{\prime }2} \\  \\
\widehat{T}_4=d-n\qquad ,\qquad \widehat{T}_5=\overline{d}-m
\end{array}
\end{equation}
where the operators in the r.h.s. of relations are defined as in (\ref{j}), (%
\ref{l}), and $\delta _1^{^{\prime }2},\delta _2^{^{\prime }2}$ are
renormalized parameters $\delta _1^2,\delta _2^2$.

The subspace of physical states ${\cal H}$ is then extracted by
conditions
\begin{equation}
\label{bg}\widehat{T}_i\left| \Phi _{phys}\right\rangle =0\qquad ,\qquad
i=1,...,5
\end{equation}
The imposition of the kinematical constraints yields that the physical wave
functions are homogeneous in $\lambda $ and $\overline{\lambda }$ of
bedegree ($n,m$) i.e.
\begin{equation}
\label{bh}\Phi \left( x,\alpha \lambda ,\overline{\alpha }\overline{\lambda }%
\right) =\alpha ^n\overline{\alpha }^m\Phi \left( x,\lambda ,\overline{%
\lambda }\right)
\end{equation}
From the standpoint of the intrinsic $M^{12}$ geometry these functions can
be interpreted as the special tensor fields being the scalars on Minkowski
space $R^{5,1}$ and, simultaneously the densities of weight ($n,m$) with
respect to the holomorphic transformations of $CP^3$. Requiring the fields (%
\ref{bh}) to be unambiguously defined on the manifold,
 the parameters $n$ and $m$ should be restricted to be integer.

Let us consider the space $^{\uparrow }{\cal H}^{[0]}(M^{12},m)$ of massive
positive frequency fields of the type (0,0) (i.e. the scalar fields on $%
M^{12}$). Such fields satisfy the mass-shell condition
\begin{equation}
\label{bi}\left( \Box -m^2\right) \Phi \left( x,\lambda ,\overline{\lambda }%
\right) =0
\end{equation}
and possess the Fourier decomposition
\begin{equation}
\label{bj}
\begin{array}{c}
\displaystyle{\Phi \left( x,\lambda ,\overline{\lambda }\right) =\int \frac{d%
\stackrel{\rightarrow }{p}}{p_0}e^{i(p,x)}\Phi \left( p,\lambda ,\overline{%
\lambda }\right) } \\  \\
p^2+m^2=0\quad ,\quad p_0>0
\end{array}
\end{equation}

The space $^{\uparrow }{\cal H}^{[0]}(M^{12},m)$ may be endowed with the
Poincar\'e-invariant and positive-definite inner product defined by the rule
\begin{equation}
\label{bk}\langle \Phi _1\left| \Phi _2\right\rangle =\int \frac{d\stackrel{%
\rightarrow }{p}}{p_0}\int\limits_{CP^3}\overline{\omega }\wedge \omega
\overline{\Phi }_1\Phi _2
\end{equation}
where the three-form $\omega $ is given by
\begin{equation}
\label{bl}\omega =\frac{\epsilon ^{abcd}\lambda _ad\lambda _b\wedge d\lambda
_c\wedge d\lambda _d}{\left( \overline{\lambda }p\lambda \right) ^2}
\end{equation}
Then $^{\uparrow }{\cal H}^{[0]}(M^{12},m)$ becomes the Hilbert space and,
as a result, the Poincar\'e representation acting on this space by the
generators (\ref{be}) is unitary. This representation can be readily
decomposed into the direct sum of irreducible ones by means of Laplace
operators $\bigtriangleup _1$ and $\bigtriangleup _2$. Namely, the subspace
of irreducible representation proves to be the eigenspace for both Laplace
operators. This implies the following
\begin{equation}
\label{bn}^{\uparrow }{\cal H}^{[0]}(M^{12},m)=\bigoplus\limits_{\stackrel{%
\scriptstyle{s_1,s_2=0,1,2,...}}{s_1\geq s_2}}{}^{\uparrow }{\cal H}%
_{s_1,s_2}(M^{12},m)
\end{equation}
and the spectrum of Laplace operators is given by the eigenvalues
\begin{equation}
\label{bm}
\begin{array}{c}
\delta _1^{^{\prime }2}=s_2\left( s_2+1\right) \ ,\quad \delta _2^{^{\prime
}2}=m^2s_1\left( s_1+3\right) \\
\\
s_1\geq s_2\qquad ,\qquad s_1,s_2=0,1,2,...
\end{array}
\end{equation}
Consequently, the subspace of physical states satisfying the quantum
conditions (\ref{bg}) is exactly $^{\uparrow }{\cal H}_{s_1,s_2}(M^{12},m)$.
The explicit expression for an arbitrary field from $^{\uparrow }{\cal H}%
_{s_1,s_2}(M^{12},m)$ reads
\begin{equation}
\Phi \left( p,\lambda ,\overline{\lambda }\right) =\Phi \left( p\right)
^{a_1...a_{s_1+s_2}b_1...b_{s_1-s_2}}\frac{\lambda _{a_1}...\lambda
_{a_{s_1}}\overline{\lambda }_{a_{s_1+1}}..\overline{\lambda }_{a_{s_1+s_2}}%
\overline{\lambda }_{b_1}...\overline{\lambda }_{b_{s_1-s_2}}}{\left(
\overline{\lambda }p\lambda \right) ^{s_1}}
\end{equation}
Here the spin-tensor $\Phi \left( p\right)
^{a_1...a_{s_1+s_2}b_1...b_{s_1-s_2}}$ is considered to be the $p$%
-transversal
\begin{equation}
\label{bp}p_{a_1b_1}\Phi \left( p\right)
^{a_1...a_{s_1+s_2}b_1...b_{s_1-s_2}}=0
\end{equation}
(for $s_1\neq s_2$) and its symmetry properties are described by the
following Young tableaux:
\unitlength=0.7mm
\special{em:linewidth 0.4pt}
\linethickness{0.4pt}
\begin{picture}(120.00,23.00)(00.00,122.00)
\put(30.00,140.00){\line(1,0){64.00}}
\put(94.00,140.00){\line(0,-1){8.00}}
\put(94.00,132.00){\line(-1,0){64.00}}
\put(30.00,132.00){\line(0,1){8.00}}
\put(30.00,132.00){\line(0,-1){8.00}}
\put(30.00,124.00){\line(1,0){32.00}}
\put(62.00,124.00){\line(0,1){8.00}}
\put(38.00,140.00){\line(0,-1){16.00}}
\put(54.00,140.00){\line(0,-1){16.00}}
\put(62.00,140.00){\line(0,-1){8.00}}
\put(70.00,140.00){\line(0,-1){8.00}}
\put(86.00,140.00){\line(0,-1){8.00}}
\put(34.00,136.00){\makebox(0,0)[cc]{$a_1$}}
\put(34.00,128.33){\makebox(0,0)[cc]{$b_1$}}
\put(45.80,135.92){\makebox(0,0)[cc]{. . .}}
\put(45.80,127.89){\makebox(0,0)[cc]{. . .}}
\put(77.94,136.10){\makebox(0,0)[cc]{. . . }}
\put(57.94,136.10){\makebox(0,0)[cc]{$a_n$}}
\put(57.94,128.07){\makebox(0,0)[cc]{$b_n$}}
\put(89.90,136.10){\makebox(0,0)[cc]{$a_m$}}
\put(120.00,136.00){\makebox(0,0)[cc]{$n=s_1-s_2$}}
\put(120.00,128.00){\makebox(0,0)[cc]{$m=s_1+s_2$}}
\end{picture}

The field $\Phi \left( p\right) ^{a_1...a_{s_1+s_2}b_1...b_{s_1-s_2}}$ can
be identified with the Fourier transform of spin-tensor field on Minkowski
space $R^{5,1}$. Together, mass-shell condition
\begin{equation}
\label{bq}\left( p^2+m^2\right) \Phi \left( p\right)
^{a_1...a_{s_1+s_2}b_1...b_{s_1-s_2}}=0
\end{equation}
and relation ($\ref{bp}$) constitute the full set of relativistic wave
equations for the mass-$m$, spin-$\left( s_1,s_2\right) $ field in six
dimensions. Thus the massive scalar field on $M^{12}$ generates fields of
arbitrary integer spins on Minkowski space.

In order to describe the half-integer spin representations of Poincar\'e
group consider the space $^{\uparrow }{\cal H}^{[1/2]}\left( M^{12},m\right)
$ of massive positive frequency fields with tensor type (1,0). These fields
possess the Fourier decomposition and may be endowed with the following
Hermitian inner product
\begin{equation}
\label{bt}\langle \Phi _1\left| \Phi _2\right\rangle _{1/2}=\int \frac{d%
\overrightarrow{p}}{p_0}\int\limits_{CP^3}\overline{\omega }\wedge \omega
\left( \overline{\lambda }p\lambda \right) ^{-1}\overline{\Phi }_1\Phi _2
\end{equation}
Then the decomposition of the space $^{\uparrow }{\cal H}^{[1/2]}\left(
M^{12},m\right) $ with respect to both Laplace operators reads
\begin{equation}
\label{bu}^{\uparrow }{\cal H}^{[1/2]}(M^{12},m)=\bigoplus\limits_{\stackrel{%
\scriptstyle{s_1,s_2=1/2,3/2,...}}{s_1\geq s_2}}{}^{\uparrow }{\cal H}%
_{s_1,s_2}(M^{12},m)
\end{equation}
where invariant subspaces $^{\uparrow }{\cal H}_{s_1,s_2}(M^{12},m)$ are the
eigenspaces of $\bigtriangleup _1$ and $\bigtriangleup _2$ with
eigenvalues
\begin{equation}
\label{bv}
\begin{array}{c}
\delta _1^{^{\prime }2}=\left( s_2-1/2\right) \left( s_2+3/2\right) \quad
,\qquad \delta _2^{^{\prime }2}=\left( s_1-1/2\right) \left( s_1+7/2\right)
\\
\\
s_1,s_2=1/2,3/2,...\quad ,\qquad s_1\geq s_2
\end{array}
\end{equation}
The explicit structure of an arbitrary field from $^{\uparrow }{\cal H}%
_{s_1,s_2}(M^{12},m)$ is
\begin{equation}
\label{bw}\Phi \left( p,\lambda ,\overline{\lambda }\right) =\Phi \left(
p\right) ^{a_1...a_{s_1+s_2}b_1...b_{s_1-s_2}}\frac{\lambda _{a_1}...\lambda
_{a_{s_1}}\overline{\lambda }_{a_{s_1+1}}..\overline{\lambda }_{a_{s_1+s_2}}%
\overline{\lambda }_{b_1}...\overline{\lambda }_{b_{s_1-s_2}}}{\left(
\overline{\lambda }p\lambda \right) ^{s_1}}
\end{equation}
where $\Phi \left( p\right) ^{a_1...a_{s_1+s_2}b_1...b_{s_1-s_2}}$ is the $p$%
-transversal tensor
\begin{equation}
\label{bx}p_{a_1b_1}\Phi \left( p\right)
^{a_1...a_{s_1+s_2}b_1...b_{s_1-s_2}}=0
\end{equation}
(for $s_1\neq s_2$) and its symmetry properties are described by the above
written Young tableaux. Consequently, from (\ref{bu}), (\ref{bv}) it follows
that the massive type (1,0) field on $M^{12}$ generates fields of arbitrary
half-integer spins on Minkowski space.

It is instructive to rewrite the inner product for two fields from $%
^{\uparrow }{\cal H}_{s_1,s_2}(M^{12},m)$ in terms of spin-tensors $\Phi
\left( p\right) ^{a_1...a_{s_1+s_2}b_1...b_{s_1-s_2}}$. The integration over
spinning variables may be performed with the use of the basic integral
\begin{equation}
\label{br}\int\limits_{CP^3}\overline{\omega }\wedge \omega =\frac{48i\pi ^3%
}{\left( p^2\right) {}^2}
\end{equation}
and the result is
\begin{equation}
\label{bo}\langle \Phi _1\left| \Phi _2\right\rangle =N\int \frac{d\stackrel{%
\rightarrow }{p}}{p_0}\overline{\Phi }_1\left( p\right)
^{a_1...a_{2s_1}}\Phi _2\left( p\right) _{a_1...a_{2s_1}}
\end{equation}
where
\begin{equation}
\label{bs}
\begin{array}{c}
\Phi _2\left( p\right) _{a_1...a_mb_1...b_n}= \\
\\
=\epsilon _{a_1b_1c_1d_1}...\epsilon
_{a_nb_nc_nd_n}p_{a_{n+1}c_{n+1}}...p_{a_mc_m}\Phi _2\left( p\right)
^{c_1...c_md_1...d_n}
\end{array}
\end{equation}
and $N$ is some normalization constant depending on $s_1$ and $s_2$.

\section{Conclusion}

In this paper we have suggested the model for a massive spinning particle in
six-dimensional Minkowski space as a mechanical system with
configuration space $M^{12}=R^{5,1}\times CP^3$. The Lagrangian of the model
is unambiguously constructed from the $M^{12}$ world line invariants when
the identical conservation is required for the classical counterparts of
Casimir operators. As a result, the theory is characterized by three
genuine gauge symmetries.

The model turns out to be completely solvable as it must, if it is a free
relativistic particle. The projection of the class of gauge equivalent
trajectories from $M^{12}=R^{5,1}\times CP^3$ onto $R^{5,1}$ represents the
two-dimensional cylinder surface of radius $\rho \sim \hbar $ with
generatings parallel to the particle's momenta.

Canonical quantization of the model naturally leads to the unitary
irreducible representation of Poincar\'e group. The requirement of the
existence of smooth solutions to the equations for the physical wave
functions results in quantization of the parameters entering Lagrangian or,
that is the same, in quantization of particle's spin.

It should be noted that switching on an interaction of the particle to the
inhomogeneous external field, one destroys the first class constraint
algebra of the model and the theory, thereby, becomes inconsistent, whereas
the homogeneous background is admissible. The physical cause underlying this
inconsistency is probably that the local nature of the inhomogeneous field
may contradict to the nonlocal behavior of the particle dynamical histories.
A possible method to overcome the obstruction to the interaction is to
involve the Wess-Zumino like invariant omitted in the action (\ref{as}). It
has the form%
$$
\Gamma =\rho \frac{(\overline{\lambda }_a\stackrel{.}{x}^{ab}\stackrel{.}{%
\lambda }_b)}{\left( \overline{\lambda }_a\stackrel{.}{x}^{ab}\lambda
_b\right) }+\overline{\rho }\frac{(\stackrel{.}{\overline{\lambda }}_a%
\stackrel{.}{x}^{ab}\lambda _b)}{\left( \overline{\lambda }_a\stackrel{.}{x}%
^{ab}\lambda _b\right) }
$$
As is easy to see, $\Gamma $ is invariant under the $\lambda $-rescalings up
to a total derivative only. This fact, however, does not prevent to say
about the particle's dynamics on $M^{12}$. The similar trick solves the
problem of interaction in the case of $d=4$ spinning particle \cite{c16}.

\section{Acknowledgments}

The authors would like to thank I. A. Batalin, I. V. Gorbunov, S. M.
Kuzenko, A. Yu. Segal and M. A. Vasiliev for useful discussions on various
topics related to the present research. The work is partially supported by
the European Union Commission under the grant INTAS 93-2058. S. L. L. is
supported in part by the grant RBRF 96-01-00482.

\section{Appendix. Half-spinorial formalism in six dimensions}

Our notations are as follows: capital Latin letters are used for Minkowski
space indices and small Latin letters for spinor ones. The metric is chosen
in the form: $\eta _{AB}=diag(-,+,...,+)$. The Clifford algebra of $8\times
8 $ Dirac matrices $\Gamma _A$ reads: $\left\{ \Gamma _A,\Gamma _B\right\}
=-2\eta _{AB}$. The suitable representation for $\Gamma _A$ is
\begin{equation}
\label{ap1}\Gamma _A=\left(
\begin{array}{cc}
0 & (\sigma _A)_{a
\stackrel{.}{a}} \\ (\widetilde{\sigma }_A)^{\stackrel{.}{a}a} & 0
\end{array}
\right) \ ,\quad
\begin{array}{c}
\sigma _A=\left\{ 1,\gamma _0,i\gamma _1,i\gamma _2,i\gamma _3,\gamma
_5\right\} \\
\widetilde{\sigma }_A=\left\{ 1,-\gamma _0,-i\gamma _1,-i\gamma _2,-i\gamma
_3,-\gamma _5\right\}
\end{array}
\end{equation}
where $\gamma _i,i=0,1,2,3,5$ are the ordinary Dirac matrices in four
dimensions. The charge conjugation matrix is defined as
\begin{equation}
\label{ap2}C=\Gamma _2\Gamma _4=\left(
\begin{array}{cc}
I & 0 \\
0 & \widetilde{I}
\end{array}
\right) \ ,\quad I=\widetilde{I}=\left(
\begin{array}{ccc}
\begin{array}{cc}
0 & 1 \\
-1 & 0
\end{array}
& | & 0 \\
--- & | & --- \\
0 & | &
\begin{array}{cc}
0 & 1 \\
-1 & 0
\end{array}
\end{array}
\right)
\end{equation}

The spinor representation of $SO(5,1)$ on Dirac spinors $\Psi =\left(
\begin{array}{c}
\lambda _a \\
\overline{\pi }^{\stackrel{.}{b}}
\end{array}
\right) $ is generated by
\begin{equation}
\label{ap3}
\begin{array}{c}
\Sigma _{AB}=-\frac 14\left[ \Gamma _A,\Gamma _B\right] =\left(
\begin{array}{cc}
(\sigma _{AB})_a{}^b & 0 \\
0 & (\widetilde{\sigma }_{AB})^{\stackrel{.}{a}}{}_{\stackrel{.}{b}}
\end{array}
\right) = \\
\\
=\left(
\begin{array}{cc}
-\frac 14\left( \sigma _A{}_{a\stackrel{.}{a}}\widetilde{\sigma }_B{}^{%
\stackrel{.}{a}b}-\sigma _B{}_{a\stackrel{.}{a}}\widetilde{\sigma }_A{}^{%
\stackrel{.}{a}b}\right) & 0 \\
0 & -\frac 14\left( \widetilde{\sigma }_A{}^{\stackrel{.}{a}b}\sigma _B{}_{b%
\stackrel{.}{b}}-\widetilde{\sigma }_B{}^{\stackrel{.}{a}b}\sigma _A{}_{b%
\stackrel{.}{b}}\right)
\end{array}
\right)
\end{array}
\end{equation}
The representation is decomposed into two irreducible ones corresponding to
the left- and right-handed Weyl spinors. It turns out that the
representation (\ref{ap3}) and its complex conjugated are equivalent: ($%
\sigma _{AB}^{*})_{\stackrel{.}{a}}{}^{\stackrel{.}{b}}=I_{\stackrel{.}{a}%
}{}^a(\sigma _{AB})_a{}^bI_b{}^{\stackrel{.}{b}},\ (\widetilde{\sigma }%
_{AB}^{*})^a{}_b=\widetilde{I}{}^a{}_{\stackrel{.}{a}}(\widetilde{\sigma }%
_{AB})^{\stackrel{,}{a}}{}_{\stackrel{.}{b}}\widetilde{I}{}^{\stackrel{.}{b}%
}{}_b$. So, one can convert the dotted spinor indices into undotted ones%
$$
\overline{\lambda }_a=I_a{}^{\stackrel{.}{a}}\stackrel{*}{\lambda }_{%
\stackrel{.}{a}}\quad ,\qquad \overline{\pi }^a=\widetilde{I}{}^a{}_{%
\stackrel{.}{a}}\stackrel{*}{\pi }{}^{\stackrel{.}{a}}
$$
While the gradient and contragradient representations are inequivalent
because of absence of an object raising and/or lowering spinor indices as
distinguished from the four-dimensional case. It is convenient to turn from
the matrices ($\sigma _A)_{a\stackrel{.}{a}},(\widetilde{\sigma }_A)^{%
\stackrel{.}{a}a}$ to ($\sigma _A)_{ab}=(\sigma _A)_{a\stackrel{.}{a}}%
\widetilde{I}{}^{\stackrel{.}{a}}{}_b,(\widetilde{\sigma }_A)^{ab}=%
\widetilde{I}{}^a{}_{\stackrel{.}{a}}(\widetilde{\sigma }_A)^{\stackrel{.}{a}%
a}$. They possess a number of relations
\begin{equation}
\label{ap4}
\begin{array}{c}
\begin{array}{cc}
(\sigma _A)_{ab}=-(\sigma _A)_{ba}{} & {}(\widetilde{\sigma }_A)^{ab}=-(%
\widetilde{\sigma }_A)^{ba}
\end{array}
\\
\\
\begin{array}{cc}
(\sigma _A)_{ab}{}(\sigma ^A)_{cd}=-2\epsilon _{abcd}{} & {}(\widetilde{%
\sigma }_A)^{ab}(\widetilde{\sigma }^A)^{cd}=-2\epsilon ^{abcd}
\end{array}
\\
\\
\begin{array}{cc}
(\sigma _A)_{ab}=-\frac 12\epsilon _{abcd}(\widetilde{\sigma }_A)^{cd}{} &
{}(\widetilde{\sigma }_A)^{ab}=-\frac 12\epsilon ^{abcd}(\sigma ^A)_{cd}
\end{array}
\\
\\
(\sigma _A)_{ab}(
\widetilde{\sigma }^A)^{cd}=2\left( \delta _a{}^c\delta _b{}^d-\delta
_a{}^d\delta _b{}^c\right) \ ,\quad (\sigma _A)_{ab}(\widetilde{\sigma }%
_B)^{ba}=-4\eta _{AB} \\  \\
(\sigma _A)_{ab}(
\widetilde{\sigma }_B)^{bc}+(\sigma _B)_{ab}(\widetilde{\sigma }%
_A)^{bc}=-2\eta _{AB}\delta _a{}^c \\  \\
(\widetilde{\sigma }_A)^{ab}(\sigma _B)_{bc}+(\widetilde{\sigma }%
_B)^{ab}(\sigma _A)_{bc}=-2\eta _{AB}\delta ^a{}_c
\end{array}
\end{equation}
Here we introduced two invariant tensors $\epsilon _{abcd}$ and $\epsilon
^{abcd}$, totally antisymmetric in indices and $\epsilon _{1234}=\epsilon
^{1234}=1$. With the aid of introduced objects one may convert the vector
indices into antisymmetric pairs of spinor ones. E.g. for a given vector $%
p_A$%
\begin{equation}
\label{ap5}p_A\rightarrow p_{ab}=p_A(\sigma ^A)_{ab}\ ,\quad p^{ab}=p_A(%
\widetilde{\sigma }^A)^{ab}\ ,\quad p_A=-\frac 14p_{ab}(\widetilde{\sigma }%
_A)^{ba}=-\frac 14p^{ab}(\sigma _A)_{ba}
\end{equation}

Consider two objects
\begin{equation}
\label{ap6}\left( \sigma _{ABC}\right) _{ab}=\frac 14(\sigma _A\widetilde{%
\sigma }_B\sigma _C-\sigma _A\widetilde{\sigma }_B\sigma _C)_{ab}\ ,\quad (%
\widetilde{\sigma }_{ABC})^{ab}=\frac 14(\widetilde{\sigma }_A\sigma _B%
\widetilde{\sigma }_C-\widetilde{\sigma }_A\sigma _B\widetilde{\sigma }%
_C)^{ab}
\end{equation}
They obey the following properties:
\begin{equation}
\label{ap7}
\begin{array}{c}
\begin{array}{cc}
\left( \sigma _{ABC}\right) _{ab}=\left( \sigma _{ABC}\right) _{ba}{} & {}(%
\widetilde{\sigma }_{ABC})^{ab}=(\widetilde{\sigma }_{ABC}).^{ba}
\end{array}
\\
\\
\begin{array}{cc}
\left( \sigma _{ABC}\right) _{ab}=\frac 16\epsilon _{ABCDEF}\left( \sigma
^{DEF}\right) _{ab}{} & {}(\widetilde{\sigma }_{ABC})^{ab}=-\frac 16\epsilon
_{ABCDEF}(\widetilde{\sigma }^{DEF})^{ab}
\end{array}
\\
\\
\left( \sigma _{ABC}\right) _{ab}(
\widetilde{\sigma }^{ABC})^{cd}=6\left( \delta _a{}^c\delta _b{}^d+\delta
_a{}^d\delta _b{}^c\right) \\  \\
\left( \sigma _{ABC}\right) _{ab}\left( \sigma ^{ABC}\right) _{cd}=(
\widetilde{\sigma }_{ABC})^{ab}(\widetilde{\sigma }^{ABC})^{cd}=0 \\  \\
\left( \sigma _{ABC}\right) _{ab}(
\widetilde{\sigma }^{DEF})^{ba}=\epsilon _{ABC}{}^{DEF}+\delta _A^{[D}\delta
_B^E\delta _C^{F]} \\  \\
(\widetilde{\sigma }_{ABC})^{ab}\left( \sigma ^{DEF}\right) _{ba}=-\epsilon
_{ABC}{}^{DEF}+\delta _A^{[D}\delta _B^E\delta _C^{F]}
\end{array}
\end{equation}
The brackets around the indices mean antisymmetrization. With the aid of
introduced objects any antisymmetric Lorentz tensor of the third rank may be
converted into a pair of symmetric bispinors.
\begin{equation}
\label{ap8}
\begin{array}{c}
M_{ABC}=\frac 1{12}(M^{ab}\left( \sigma _{ABC}\right) _{ba}+M_{ab}(
\widetilde{\sigma }_{ABC})^{ba}) \\  \\
M^{ab}=M^{ABC}(\widetilde{\sigma }_{ABC})^{ab}\ ,\quad M_{ab}=M^{ABC}\left(
\sigma _{ABC}\right) _{ab}
\end{array}
\end{equation}
In conclusion we write out the Fierz identities and rules of complex
conjugation for different spinor bilinears. For the sake of simplicity
we omit the contracted spinor indices throughout this paper, e. g. ($\chi
\widetilde{\sigma }_A\psi )=\chi _a(\widetilde{\sigma }_A)^{ab}\psi _b,(\chi
\widetilde{\sigma }_{ABC}\psi )=\chi _a(\widetilde{\sigma }_{ABC})^{ab}\psi
_b$%
\begin{equation}
\label{ap9}
\begin{array}{c}
\psi _a\chi _b=\frac 14(\psi
\widetilde{\sigma }_A\chi )\sigma ^A{}_{ab}+\frac 1{12}(\psi \widetilde{%
\sigma }_{ABC}\chi )\left( \sigma ^{ABC}\right) _{ab} \\  \\
\chi ^b\psi _a=\frac 14\left( \chi \psi \right) \delta _a{}^b-\frac 12\left(
\chi \sigma _{AB}\psi \right) \left( \sigma ^{AB}\right) _a{}^b \\
\\
\left( \psi \chi \right) ^{*}=\left(
\overline{\psi }\overline{\chi }\right) \quad ,\qquad \left( \psi \overline{%
\chi }\right) ^{*}=-\left( \overline{\psi }\chi \right) \\  \\
(\chi
\widetilde{\sigma }_A\psi )^{*}=(\overline{\chi }\widetilde{\sigma }_A%
\overline{\psi })\quad ,\qquad (\overline{\chi }\widetilde{\sigma }_A\psi
)^{*}=-(\chi \widetilde{\sigma }_A\overline{\psi }) \\  \\
(\overline{\chi }\widetilde{\sigma }_{ABC}\psi )^{*}=-(\chi \widetilde{%
\sigma }_{ABC}\overline{\psi })\ ,\quad (\chi \widetilde{\sigma }_{ABC}\psi
)^{*}=(\overline{\chi }\widetilde{\sigma }_{ABC}\overline{\psi })
\end{array}
\end{equation}
Analogous relations take place for spinor with upper indices.

\end{document}